\documentclass[sigconf,screen]{acmart}
\copyrightyear{2025}
\acmYear{2025}
\setcopyright{cc}
\setcctype{by}
\acmConference[PROMISE '25]{21st International Conference on Predictive
Models and Data Analytics in Software Engineering}{June 26, 2025}{Trondheim,
Norway}
\acmBooktitle{21st International Conference on Predictive Models and Data
Analytics in Software Engineering (PROMISE '25), June 26, 2025, Trondheim,
Norway}\acmDOI{10.1145/3727582.3728682}
\acmISBN{979-8-4007-1594-5/2025/06}
\usepackage{graphicx} 
\usepackage{caption}
\usepackage{colortbl}
\usepackage{multirow}
\usepackage{array}
\usepackage{flushend}
\usepackage{tablefootnote}
\usepackage{hyperref}
\usepackage{balance}
\usepackage{todonotes}
\usepackage{url}
\usepackage{booktabs}  
\usepackage{pifont}    

\AtBeginDocument{%
  }

\begin{document}

\title{LO2: Microservice API Anomaly Dataset of Logs and Metrics}
\author{Alexander Bakhtin}
\affiliation{%
  \institution{University of Oulu}
  \country{Finland}
}
\email{alexander.bakhtin@oulu.fi}

\author{Jesse Nyyssölä}
\affiliation{%
  \institution{University of Helsinki}
  \country{Finland}
}
\email{jesse.nyyssola@helsinki.fi}

\author{Yuqing Wang}
\affiliation{%
  \institution{University of Helsinki}
  \country{Finland}
}
\email{yuqing.wang@helsinki.fi}

\author{Noman Ahmad}
\affiliation{%
  \institution{University of Oulu}
  \country{Finland}
}
\email{noman.ahmad@oulu.fi}

\author{Ke Ping}
\affiliation{%
  \institution{University of Helsinki}
  \country{Finland}
}
\email{ke.ping@helsinki.fi}

\author{Matteo Esposito}
\affiliation{%
  \institution{University of Oulu}
  \country{Finland}
}
\email{matteo.esposito@oulu.fi}

\author{Mika Mäntylä}
\affiliation{%
  \institution{University of Helsinki}
  \country{Finland}
}
\email{mika.mantyla@helsinki.fi}

\author{Davide Taibi}
\affiliation{%
  \institution{University of Oulu}
  \country{Finland}
}
\email{davide.taibi@oulu.fi}
\renewcommand{\shortauthors}{Bakhtin et al.}

\begin{abstract}
\textbf{Context}. Microservice-based systems have gained significant attention over the past years.
A critical factor for understanding and analyzing the behavior of these systems is the collection of monitoring data such as logs, metrics, and traces.
These data modalities can be used for anomaly detection and root cause analysis of failures.
In particular, multi-modal methods utilizing several types of this data at once have gained traction in the research community since these three modalities capture different dimensions of system behavior.

\noindent\textbf{Aim}. We provide a dataset that supports research on anomaly detection and architectural degradation in microservice systems. We generate a comprehensive dataset of logs, metrics, and traces from a production microservice system to enable the exploration of multi-modal fusion methods that integrate multiple data modalities. 

\noindent\textbf{Method}. We dynamically tested the various APIs of the MS-based system, implementing the OAuth2.0 protocol using the Locust tool. For each execution of the prepared test suite, we collect logs and performance metrics for correct and erroneous calls with data labeled according to the error triggered during the call.

\noindent\textbf{Contributions}. 
We collected approximately 657,000 individual log files, totaling over two billion log lines. In addition, we collected more than 45 million individual metric files that contain 485 unique metrics.
We provide an initial analysis of logs, identify key metrics through PCA, and discuss challenges in collecting traces for this system. Moreover, we highlight the possibilities for making a more fine-grained version of the data set. This work advances anomaly detection in microservice systems using multiple data sources. 
\end{abstract}

\ccsdesc[500]{Computer systems organization~Distributed architectures}
\ccsdesc[500]{Security and privacy~Intrusion/anomaly detection and malware mitigation}
\ccsdesc[500]{Information systems~Test collections}
\newcounter{excerpt}
\keywords{Microservice Systems, Anomaly Detection, OAuth2.0, Logs, Traces, Metrics, Multi-modal Fusion Methods,  Dataset}

\maketitle

\section{Introduction}
\label{sec:Intro}
We live in a constantly connected world where Microservice-based systems (MSS) permeate our daily activities. Due to their pervasiveness, industry and researchers are continually investigating multiple aspects, including architectures, patterns, antipatterns, performances, and the evolution of MSS \cite{taibi2017processes,bakhtin2025network, schneider2023automatic, amadini2024pick}.

In MSS research and development,  we must execute complex and large systems to collect tracing data to analyze their behavior, gather insights, and test new and emerging ideas. Nonetheless, executing and collecting relevant nontrivial data is a lengthy process that requires considerable effort. Despite the daunting need for public and real-world-like datasets, researchers investigated a few related ones in the literature and executed similar systems every time. 

A particular domain where microservice (MS) execution is needed is anomaly detection. A microservice anomaly generally refers to unexpected behaviors or performance issues within a microservices architecture that deviates from normal behavior \cite{8957683,nist800-190}. Anomalies can manifest as increased latency, resource exhaustion, unexpected failures, or performance degradation, often resulting from complex interactions within the distributed system \cite{zhang2022deeptralog, yu2023nezha}.

These anomalies can be generated through various reliability testing practices. For example, in fault injection testing, the system is introduced with faults such as network issues, hardware exhaustion, system crashes, and configuration errors \cite{chaosengineering2024}. Other practices include stress testing and endurance testing, which refer to running the system under a heavy load or long duration, respectively. Current research has placed emphasis on such performance-related anomalies (See Chapter \ref{sec:related}).

Traces, logs,  and metrics are fundamental for anomaly detection in MSS \cite{yu2023nezha}.
Anomaly detection requires the availability of large datasets containing data from different sources, including the log or traces of each service individually or metrics describing the performance of the infrastructure or an individual service container. 
 Traces capture the flow of requests as they move through service instances (i.e., spans) within the MSS. Logs record execution details of service instances. Metrics are quantitative measurements that track the performance of an entity, such as MS, or the entire host system over time.

Although MSS datasets that include all three modalities have been openly published \cite{yu2023nezha, LeeYCSL23, AIOpsChallenge2021}, most of them either provide data on OSS benchmark (\cite{yu2023nezha, LeeYCSL23, li2022actionable}) or proprietary, closed systems (\cite{li2022actionable, AIOpsChallenge2021}. Other datasets do not provide all three modalities \cite{zhang2022deeptralog, li2022actionable}.


To help researchers merge the gap above, we aimed to provide a dataset of logs, metrics, and traces based on executing commercial and production yet open-sourced MS-based system, Light-OAuth2.
In this context, our work contributes to the community by providing:
   \noindent\textbf{A dataset of monitoring data for a production-ready (non-demo) MSS.}
   Many existing similar datasets use either benchmark demo systems such as \emph{TrainTicket} or \emph{DeathStarBench}, or commercial proprietary MSS whose code cannot be examined. We leverage the OSS production system LightOAuth2, as such approach allows, on the one hand, to use a \emph{"realistic"} system, but on the other hand, have its code available for static analysis and anomaly injection to deepen the potential analysis of the system. We set up the data collection process for all three modalities, i.e., \textbf{logs}, \textbf{metrics}, and \textbf{traces}, but in the case of LightOAuth2, we only obtained a significant amount of logs and metrics.
   
    \noindent The dataset is prepared and shared according to FAIR principles \cite{force11_fair}: Findable, Accessible, Interoperable, and Re-usable.
   
   \noindent\textbf{Scripts and tools to replicate our study.}
   We share a package \cite{replication}, which provides several scripts and detailed instructions on how to run them, enabling an end-to-end process of deploying the system, executing the tests, gathering the data, and saving it with appropriate structure. 
   
   \noindent\textbf{Preliminary analysis.}
    For the logs, we perform anomaly detection with binary labels using different representations of log data and analysis models. Additionally, we analyze how well each of the services' log data performs in detecting each error introduced in the data collection process and provide the F-score for all such combinations.
   For the metrics, we identify the most significant host metrics by performing Principal Component Analysis. We discuss the issues that arose during the data collection process for the traces.
   
   \noindent\textbf{Ideas for future use of the dataset.}
   We discuss a few applications of the dataset, including anomaly detection in microservices using multi-modal data.

\textbf{Paper Structure:}  Section~\ref{sec:related} discussed related datasets, Section~\ref{sec:DataCollection} describes the data collection process while Section~\ref{sec:DataDescription} describes the collected data structure, and Section~\ref{sec:PreliminaryAnalysis} provide a preliminary analysis of the data; Section~\ref {sec:future} highlights possible future use, and Section~\ref{sec:udaptes} explains how the community can contribute to updating the dataset; Section \ref{sec:howto} explains how to access the dataset and Section~\ref{sec:license} presents the license statement; Section~\ref{sec:Threats} discusses the threats to the validity of our study. Finally, Section~\ref{sec:Conclusion} concludes this work.

\section{Related Work}
\label{sec:related}
Our novel dataset improves upon existing uni-modal \cite{LeeYCSL23}, bi-modal \cite{zhang2022deeptralog} and multi-modal \cite{yu2023nezha, li2022actionable, AIOpsChallenge2021} datasets by providing monitoring data of an OSS (as opposed to proprietary \cite{li2022actionable, AIOpsChallenge2021}) and production (as opposed to benchmark \cite{yu2023nezha, zhang2022deeptralog, LeeYCSL23, rcaeval}) microservice system.

Yu et al. \cite{yu2023nezha} provide a dataset \cite{Nezha} to evaluate their \emph{Nezha} framework for Root Cause Analysis (RCA).
The authors provide the evaluation dataset of logs, metrics, and traces mined from \emph{TrainTicket} and \emph{OnlineBoutique}, part of \emph{DeathStarBench} (DS) family of OSS MSS. The authors inject several performance issues to the container deployment, such as CPU contention, CPU consumption, and Network delay, as well as code defect injections for Java and Python languages to emulate returning errors and exceptions.

Zhang et al. \cite{zhang2022deeptralog} propose the \emph{DeepTraLog} deep learning approach for anomaly detection (AD) using logs and traces of MSS. To evaluate the framework, they provide the \emph{DeepTraLog} dataset \cite{DeepTraLog} of logs and traces mined from the \emph{TrainTicket} MSS. The authors use the main implementation as well as an additional 14 versions of the system with different implemented faults provided by developers of \emph{TrainTicket}.

Lee et al. \cite{LeeYCSL23} propose the \emph{Eadro} framework to integrate anomaly detection and root cause analysis from logs, traces, and metrics (referred to in the paper as KPIs) using Deep Learning. The authors share the evaluation dataset of logs, traces, and metrics \cite{Eadro} gathered from \emph{TrainTicket} and \emph{DeathStarBench-SocialNetwork} systems. Three performance anomalies are injected into the systems, namely CPU exhaustion, network delay, and packet loss.

Li et al. \cite{li2022actionable} propose the \emph{D\'ej\`aVu} framework to perform root cause analysis of recurring failures in MSS. The approach is based on metrics only, and authors share the evaluation dataset \cite{Dejavu}. The authors evaluate their approach using real and injected faults on \emph{TrainTicket} as well as three proprietary MSS: commercial ISP MSS, commercial bank MSS, and its Oracle database. For the three MSS, performance faults such as CPU exhaustion, network delay, packet loss, database limit, low memory, and pod failure. For the Oracle database, the authors manually labeled real historical failures in different classes.

Researchers from Tsinghua University provided the \emph{AIOps challenge} dataset \cite{AIOpsChallenge2021} for the challenge of the same name in 2021. The dataset contains all three modalities: logs, metrics, and traces. The challenge in 2021 was to perform Anomaly Detection on this multi-modal data. The injected anomalies include packet loss, high memory usage, network delay, high disk space usage, CPU exhaustion, and JVM resource exhaustion.

Pham et al. provide the RCAEval framework \cite{rcaeval}, which includes multimodal data of logs, metrics and traces for three OSS microservice benchmarks. The authors injected performance and network anomalies to the container deployment as well introduced code defects into the source code of the services.

The summary of the related datasets is provided in Table \ref{tab:datasets}.

\begin{table*}[t]
\centering
\renewcommand{\arraystretch}{1.35}

\caption{Existing MSS Monitoring Datasets}
\label{tab:datasets}
\resizebox{\linewidth}{!}{%
\begin{tabular}{l p{0.35\linewidth} ccc p{0.46\linewidth} p{0.35\linewidth} l}
\hline
\multirow{2.3}{*}{\textbf{Dataset}} & \multirow{2.3}{*}{\textbf{MSS}} & \multicolumn{3}{c}{\textbf{Monitoring data}} & \multirow{2.3}{*}{\textbf{Anomaly type }} &  Injection method &\multirow{2.3}{*}{\textbf{Goal}} \\ 
\cmidrule(lr){3-5}
 & & \textbf{Logs} & \textbf{Metrics} & \textbf{Traces} &  \\ 
\hline
Nezha \cite{Nezha, yu2023nezha} & TrainTicket, DS-OnlineBoutique & \checkmark & \checkmark & \checkmark & Performance (4), 
Code defects (1) & Performance: injection method not described, Code defects: Customized fault injectors  & RCA\\ 
\multirow{2}{*}{DeepTraLog \cite{DeepTraLog, zhang2022deeptralog}} & \multirow{2}{*}{TrainTicket} &  \multirow{2}{*}{\checkmark} &  &  \multirow{2}{*}{\checkmark} & Asynchronous interaction (3), Multi-Instance interaction (3), Configuration (4), Code defects (4) & Not described &\multirow{2}{*}{AD} \\ 
Eadro \cite{Eadro, LeeYCSL23} & TrainTicket, DS-SocialNetwork & \checkmark & \checkmark & \checkmark & Performance (3) 
& Chaosblade simulation & AD+RCA \\ 
\multirow{2}{*}{D\'ej\`aVu \cite{Dejavu,li2022actionable}}& TrainTicket, Commercial ISP MSS, Commercial bank MSS + Oracle database &   &   \multirow{2}{*}{\checkmark}&  & \multirow{2}{*}{Performance (10),  Functional anomalies/Fail-stop (3)} & Failure
injection at random locations with ChaosMesh, manual injection
& \multirow{2}{*}{RCA} \\ 
AIOps Challenge 2021 \cite{AIOpsChallenge2021} & 2xCommercial bank MSS & \checkmark & \checkmark & \checkmark &  Performance (7) 
& Not described & AD, RCA \\ 
RCAEval \cite{rcaeval} & Train-Ticket, DS-OnlineBoutique, SockShop & \checkmark & \checkmark & \checkmark &  Performance (6), Code Defects (5) &Failure injection, Code modification& RCA \\
\hline 
Our work & Production OSS MSS&  \checkmark &\checkmark&&API errors& Negative testing& AD \\
\hline
\end{tabular}%
}
\end{table*}

\section{Data Collection}
\label{sec:DataCollection}
In this section, we describe the studied system, the testing framework, running the system with the tests, and querying and collecting all data.
\subsection{The System Under Study: Light-OAuth2}
The Light-OAuth2 system\footnote{\url{https://doc.networknt.com/getting-started/light-oauth2/}} is an Open-Source implementation of the OAuth2.0 authorization protocol \cite{rfc6749} developed by \texttt{networknt}. It consists of 7 microservices and a MySQL database, although other database options are also supported. 
The selected project was still under development at the time of selection and test preparation. 
After the commencement of data collection, the project has been archived\footnote{\url{https://github.com/networknt/light-oauth2}}.
Table \ref{tab:services_short} shows the list and descriptions of all services.
\begin{table}[t]
\renewcommand{\arraystretch}{1.5} 
\caption{List of Services of Light-OAuth2 System and Implemented Locust Tasks}
\label{tab:services_short}
\centering

\resizebox{0.9\linewidth}{!}{%
\begin{tabular}{cp{0.15\linewidth}cc}
\hline
\multicolumn{1}{c}{\textbf{Service}} & \multicolumn{1}{c}{\textbf{Function}}      & \multicolumn{1}{c}{\textbf{Tasks (correct)}} & \multicolumn{1}{c}{\textbf{Tasks (errors)}} \\ \hline
oauth2-token                         & OAuth2.0                & 3\                                            & 12 \\
oauth2-code                          & OAuth2.0                        & 2                                          & 9                                        \\
oauth2-service                       & CRUD          & 5                                            & 8                                           \\
oauth2-client                        & CRUD           & 5                                            & 10                                          \\
oauth2-user                          & CRUD          & 6                                            & 11                                          \\
oauth2-refresh-token                 & CRUD  & 3                                            & 3                                           \\
oauth2-key                           & Publ. Keys       & 0                                            & 0                                           \\
mysqldb                              & Database              & -                                            & -                                           \\ \hline
\end{tabular}%
}

\end{table}

\subsection{Test preparation}
The goal of the created test suite is to provide broad coverage of the Light-OAuth2 APIs. In particular, we considered all standard OAuth2.0 flows within this framework as well as the management of entities related to these flows, such as clients, services, and users.
Our tests cover the following flows (Grant Types) of the OAuth2.0 protocol:
\begin{itemize}
    \item Authorization code (without PKCE extension)
    \item Authorization code (with PKCE extension)
    \item Client credentials
    \item Refresh token
\end{itemize}
Additionally, we implemented Create, Read, Update, and Delete operations for Client, User, and Service registration services.

We excluded the outdated and insecure Resource Owner Password flow\footnote{\url{https://auth0.com/docs/get-started/authentication-and-authorization-flow/resource-owner-password-flow}}, as well as custom functionalities of Light-OAuth2 implementation, such as mapping of clients to services and management of authentications keys. Neither do we consider the scopes of the issued OAuth2.0 tokens, since we do not simulate an actual system utilizing OAuth2.0.

Tests are prepared and executed with Locust\footnote{\url{https://locust.io}}, an Open-Source load-testing tool written in Python that allows the testing RESTful APIs and easily defining complex scenarios mimicking different use cases and business logic of the APIs. Each function calling an API is referred to as \emph{task}. A collection of tasks executed together is called a \emph{test}.

Among the authors, the two with the most relevant experience in API quality and security analysis
studied the Light-OAuth2 API documentation to implement the relevant tests in Locust. Moreover, these authors reviewed each others' test suites to ensure compliance with the aforementioned documentation.

In the case of CRUD operations, the correct requests to the API were implemented, as well as all possible errors in the request parameters as described in the documentation. As for the OAuth2.0 flows, both the Light-OAuth2 documentation and the official specification of the OAuth2.0 protocol were used, allowing us to implement these tests in a way that is portable to another OAuth2.0 implementation while avoiding ad-hoc additions of Light-OAuth2.

We split the Locust tasks into two groups - those implementing \emph{correct} requests according to the documentation of each service and those introducing \emph{errors}, aka negative tests \cite{hora2024test}, in the requests on purpose for each error described in the documentation\footnote{\url{https://doc.networknt.com/service/oauth/service/}}. For example, the Service Registration MS \footnote{\url{https://doc.networknt.com/service/oauth/service/service/}} contains the endpoint \texttt{/oauth2/service} which accepts GET requests to query an existing service, POST requests to register a new service, PUT requests to update a service, and DELETE requests to delete a service. All return status code 200 in case of a properly formulated request, and the corresponding tasks are labeled as \emph{correct}. In case a provided \texttt{serviceId} does not exists, GET, PUT and DELETE requests will return status code 404 \texttt{SERVICE\_NOT\_FOUND}; in case a given \texttt{serviceId} already exists in the database, the POST request will return status code 400 \texttt{SERVICE\_ID\_EXISTS}. All of these four API errors have a corresponding Locust error task, and are labeled as \emph{error} tests in the dataset, with each error having its own label.

Table \ref{tab:services_short} shows the number of tasks implemented for each service. 
\subsection{Test execution}
To execute the tests and collect the data for future analysis, we perform several \emph{runs} of the test suite on the studied system.

We define a \emph{run} as a full execution of the following procedure:

\begin{enumerate}
    \item deploy the Docker Compose of the system,
    \item wait for all services and MySQL to become operational,
    \item run only the \emph{correct locust tasks} for 60s (\emph{correct test}),
    \item query logs, traces, and metrics,
    \item for each \emph{error locust task}:
    \begin{enumerate}
    \item run correct + that particular error task for 10s (\emph{error test}),
    \item query logs, traces, and metrics,
    \end{enumerate}
    \item stop Docker Compose.
\end{enumerate}

The system was deployed, and tests were executed hourly on our research server as a \texttt{cronjob} in July-August 2024 until the disk space reserved on the server was filled up (540 GB).
We did \textbf{not attempt fault injection}, such as hardware/host anomalies, network delays or slow disk usage, but instead focused on the anomalies that occur in the system due to erroneous API usage in this work.

We produced a total of 1740 \emph{runs} consisting of 54 \emph{tests} in each run (1 with \emph{correct tasks} only + 53 introducing one \emph{error task} each). Uncompressed raw data occupies about 540 GB.

\subsection{Data collection}

Container logs are obtained by querying \texttt{docker logs} with the corresponding container ID.
We also configured Locust to export its logs during each run, including statements for each query done to the MSS.

Traces are obtained by injecting the \texttt{jaeger-agent}\footnote{\url{https://www.jaegertracing.io/}} into the Docker Compose deployment. Metrics are obtained similarly by injecting Prometheus' node exporter\footnote{\url{https://prometheus.io/}} into the deployment. To collect the data, we utilized the API of the respective systems to fetch the data every five seconds and store it into JSON files for the metrics and CSV files for the traces. 

Example of collected log, metrics and trace data are available in Listings \ref{tab:sample_logs}, \ref{tab:sample_metrics}, and \ref{tab:sample_traces}.

\section{Data Description}
\label{sec:DataDescription}
\begin{figure*}[t]
    \centering
    \includegraphics[width=0.8\linewidth]{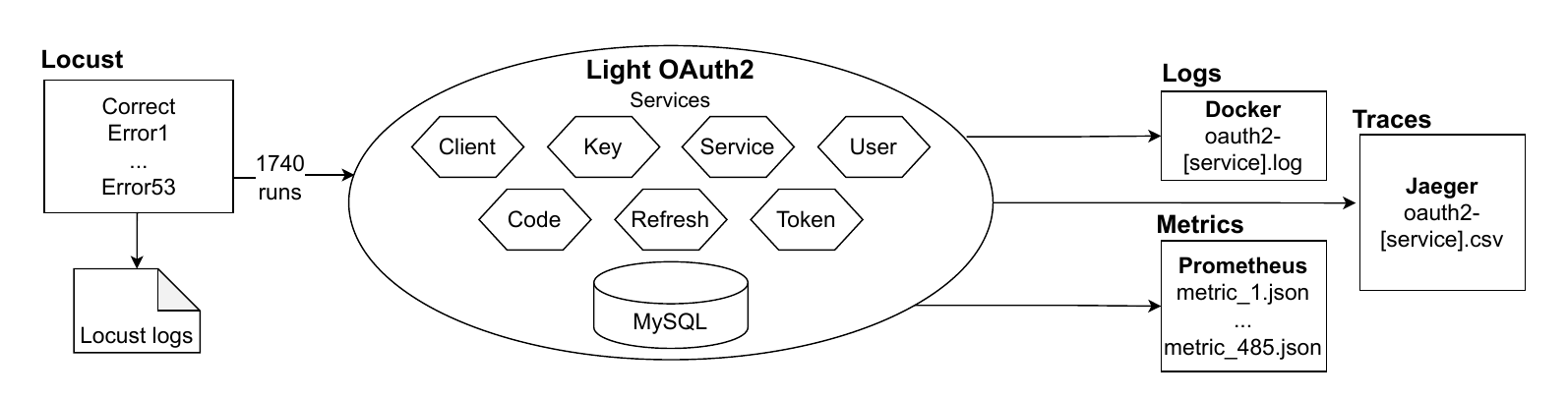}
    \caption{Data Collection Process}
    \label{fig:dcp}
\end{figure*}
\subsection{Logs}
\label{sec:desc_logs}
\begin{figure}[t]
    \centering
    \includegraphics[width=0.9\linewidth]{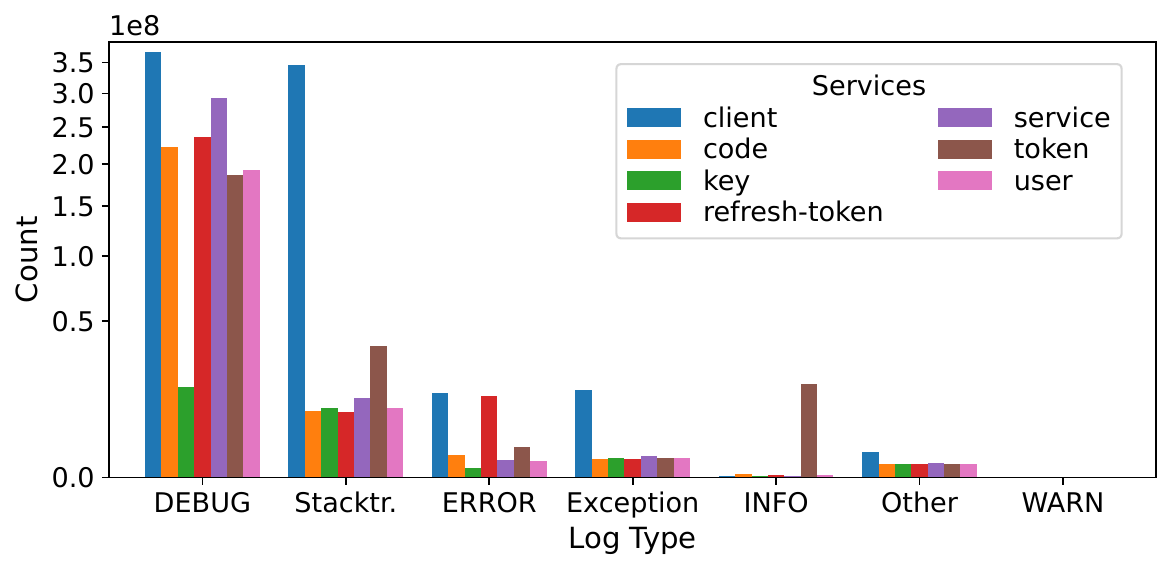}
    \caption{Number of Log Lines per Type and Service}
    \label{fig:logtypes}
\end{figure}

The log files are generated by each container, which corresponds to an individual service. Files are saved for each run and each test. There are approximately 657,000 individual log files for the services (1740 runs * 54 tests  * 7 services). The content of the log files comes directly from the logger used in each system at the logging level \texttt{DEBUG}. The service logs contain over \textbf{two billion} log lines, as illustrated in Figure~\ref{fig:logtypes}. In the figure, capital letters refer to an explicit logging level in the log line, while lowercase log types are identified with regular expressions. Note that the figure uses square root scaling on the y-axis. 
\subsection{Metrics}
\label{sec:desc_metrics}
Prometheus metrics are exported as JSON files. In total, there are 485 metrics, which means that across all runs and tests, there are over 45.5 million individual metric files. This approach provides a granular data structure, but for practical purposes, it is sensible to combine several files into a longer time series. We provide data processing scripts for this purpose \cite{dataset_repo}. The format of individual metrics varies as they can have different parameters, but they generally contain timestamp-value pairs that effectively form the time series.

\subsection{Traces}
\label{sec:desc_traces}
Traces are exported as CSV files. Similarly to log files, each combination of run, test, and service has its own CSV file with the following fields: Trace ID, Span ID, Operation Name, Start Time, End Time, and Tags. However, in our current dataset, each trace currently contains only a single span. The Light-OAuth2.0 system only includes requests that make direct calls from the system to the MySQL service, while this system was not implemented initially to provide traceability across sub-operations within services. For example, although a single trace might capture a high-level request for the MySQL service, it did not record finer-grained operations, such as individual POST and GET requests, parameter parsing, authentication checks, or response processing. 
To achieve such a level of detail, we would need to modify Light-OAuth2.0’s source code to manually instrument sub-operations within each service interaction using OpenTelemetry. That requires significant code changes and restructuring, which is infeasible due to our limited resources and the system’s archived status. Furthermore, most services do not perform database operations that the tracing agent can catch so the respective CSVs will be empty.

\subsection{Metadata}
\label{sec:desc_meta}
In addition to the data, we also provide the metadata. Each run has a log file containing the response logs made by the Locust tests. These are not part of the Light-OAuth2 system but can help in analyzing the results. For instance, if there is an error but the log message given by the system does not provide much information, the Locust log shows exactly which locust task caused it. It also has timestamps, which can be used as a ground truth of when each request was sent. 

In our data repository \cite{dataset_repo} we provide data appendix that contains statistics of the whole dataset such as file size distributions. Studies have shown that simple predictors such as sequence length can achieve competitive anomaly detection rates on certain datasets \cite{criticalreviewlog}. As such, file size distributions too can provide a valuable and computationally efficient predictor in anomaly detection.

\section{Preliminary Analysis}
\label{sec:PreliminaryAnalysis}
Our preliminary analysis will focus on analyzing the software execution through log files. Additionally, we perform Principal Component Analysis of the metrics' time series.
\subsection{Log analysis}
For the analysis of the log files, we use an existing tool for benchmarking log anomaly detection algorithms called LogLead \footnote{\url{https://github.com/EvoTestOps/LogLead}}\cite{Mantyla_LogLead_-_Fast_2024}. It supports several representation methods for the data and machine learning models for anomaly detection. In our case, we want to learn to identify the error tests from the correct ones. To that end, we represent each log file (i.e., a combination of run, test, and service) as their own sequence, and we can label them based on the error tag on the test level.

We perform the log analysis in two stages. In the first stage, we apply anomaly detection models with different representation approaches similar to the previous work \cite{nyyssola2024unsuper}. The models included in the approach are DecisionTree (DT), LogisticRegression (LR), IsolationForest (IF), KMeans, RarityModel (RM) and Out-of-Vocabulary Detector (OOVD). DT and LR are the established supervised models from the \emph{scikit-learn} library. IF and KMeans are the unsupervised models from the same library. RM and OOVD are efficient custom unsupervised models introduced in \cite{nyyssola2024unsuper}. The unsupervised models work based on the idea that we only use correct training data and measure the deviation. The representations for the log lines are words, trigrams, and parsed event IDs. 

The approach in the first stage does not distinguish between individual error types or services. However, the logs from different services vary significantly in content, making it impractical to combine them for analysis by simply appending them. To address this, in the second stage of the analysis, we apply a highly granular anomaly detection approach that analyzes each combination of error cases and service logs individually, producing a 53x7 result matrix. This approach is not practically viable as it requires knowledge of the error cases beforehand, but for the purposes of this paper, it serves as a demonstration of the data and what could be possible. Based on these results, we can propose directions for future work in Section \ref{sec:future}.  

Both stages use a sample of 100 runs where we select one error test and the correct test from each run. In the first stage, the error is selected randomly for each run, and we utilize the log file of each service, which gives a total of 700 sequences. In the second stage, only a single service is selected at a time, which results in 200 sequences, but the analysis is done on each service separately. In both stages, the data is evenly divided between normal and anomalous, and it is split randomly with a 50-50 ratio for training and testing. For the first stage, we run the whole setup 10 times and report the means and standard deviations. In the second stage, we run once due to the high number of error-service pairs. 

For the supervised models, we use the F1-score metric, which is the standard practice in the field of anomaly detection. For the unsupervised models, we prioritize the AUCROC, which measures the area under the receiver-operating characteristic curve as it measures the performance across all possible thresholds, and the models are very threshold-dependent. For reference, we also provide the AUCROC score for the supervised models and use an optimizer with labels to find the best possible threshold regarding F1-score for the unsupervised models.

The results for the first stage are in Table \ref{tab:log1}. The asterisk signifies that the unsupervised models used an optimizer for F1-score threshold selection. All of the optimized scores are close to 0.667 which, in practice, means that the model does not work and the optimizer defaults to assigning everything as errors. This leads to the F1-score of 0.667, as shown in the formula below.
\[
F_1 = 2 \times \frac{\left(\frac{50}{50 + 50}\right) \times \left(\frac{50}{50 + 0}\right)}{\left(\frac{50}{50 + 50}\right) + \left(\frac{50}{50 + 0}\right)} = 2 \times \frac{0.5 \times 1.0}{0.5 + 1.0} \approx 0.667 
\]
The AUCROC scores for the unsupervised models are similarly low at around 0.5, which indicates random classification performance. From all the models only DecisionTree performed reasonably with F1-scores of over 0.7. The representation selection did not have a significant impact. In summary, the results in the first stage suggest that simple concatenation of the services is not a good approach for detecting errors in the dataset. 

\begin{table*}[t]
\caption{Log Analysis Scores by Model and Representation Type}
\label{tab:log1}
\centering
\resizebox{0.85\linewidth}{!}{
\begin{tabular}{l|cc|cc|cc}
\hline
& \multicolumn{2}{c|}{\textbf{Words}} & \multicolumn{2}{c|}{\textbf{Trigrams}} & \multicolumn{2}{c}{\textbf{Event IDs}} \\
\textbf{Model} & \textbf{F1-score} & \textbf{AUCROC} & \textbf{F1-score} & \textbf{AUCROC} & \textbf{F1-score} & \textbf{AUCROC} \\ \hline
IsolationForest & 0.666* $\pm$ 0.011 & 0.489 $\pm$ 0.019 & 0.666* $\pm$ 0.007 & 0.502 $\pm$ 0.018 & 0.663* $\pm$ 0.008 & 0.479 $\pm$ 0.014 \\
KMeans & 0.674* $\pm$ 0.011 & 0.530 $\pm$ 0.018 & 0.671* $\pm$ 0.009 & 0.529 $\pm$ 0.017 & 0.666* $\pm$ 0.007 & 0.516 $\pm$ 0.025 \\
RarityModel & 0.666* $\pm$ 0.011 & 0.539 $\pm$ 0.011 & 0.666* $\pm$ 0.007 & 0.580 $\pm$ 0.015 & 0.663* $\pm$ 0.008 & 0.511 $\pm$ 0.017 \\
OOV Detector & 0.666* $\pm$ 0.011 & 0.416 $\pm$ 0.015 & 0.661* $\pm$ 0.012 & 0.533 $\pm$ 0.012 & 0.664* $\pm$ 0.017 & 0.534 $\pm$ 0.012 \\
DecisionTree & 0.720 $\pm$ 0.021 & 0.731 $\pm$ 0.019 & 0.731 $\pm$ 0.025 & 0.727 $\pm$ 0.018 & 0.700 $\pm$ 0.019 & 0.726 $\pm$ 0.018 \\
LogisticReg. & 0.467 $\pm$ 0.125 & 0.513 $\pm$ 0.020 & 0.478 $\pm$ 0.046 & 0.500 $\pm$ 0.018 & 0.508 $\pm$ 0.066 & 0.547 $\pm$ 0.027 \\
\hline
\end{tabular}%
}
\end{table*}

The whole matrix for the second stage is presented in Table \ref{tab:log2}. Due to its superior performance, we used the DecisionTree model in the analysis. \emph{Words} were selected as the log representation format as it is computationally the most efficient, and there were no significant performance gains on the other representations.

On average, the Client and Token service logs contain the most relevant content for anomaly detection, while the Key log performs the worst. The performance of other services is highly dependent on the error case being tested, but they can reach high accuracy when used for an appropriate error type. In fact, all other services except for the Key service managed to achieve perfect anomaly detection accuracy on some error types. This illustrates that having multiple individually trained models can provide comprehensive coverage across all test cases.  

The Key service is expected to be the worst predictor because we did not implement Locust tasks for it since its functionality is not part of the standard OAuth2.0 protocol (see Table \ref{tab:services_short}).

The Token service being the best predictor of the anomaly can be explained by the fact that an OAuth flow is a multi-step process, with the Token service being used at the last step, so a lot of successful operations of that service indicate that it happened during the \emph{correct test}, while at any \emph{error test}, Token service would only be invoked correctly a few times due to the errors encountered at the previous steps.

\subsection{Metric analysis}
Since the exported data contains 485 individual metrics, some containing several time series due to different parameters (see Section~\ref{sec:desc_metrics}), we perform Principal Component Analysis to identify the most relevant metrics.

From the 172 timestamps and 1124 feature vectors (we consider each metric time series as a separate feature vector but remove Prometheus' metrics), we obtain $min(172, 1124)=172$ components in the PCA. We observe that already the Top-3 components explain more than $99\%$ of data variance.

We select Top-5 contributing features from these Top-3 contributing components. We obtain a set of only ten distinct features (out of $min=5$ and $max=15$):
disk written bytes total (for the device used by Docker),
filesystem avail. bytes (for the device and filesystem used by Docker),
memory AnonHugePages bytes,
memory AnonPages bytes,
memory Committed AS bytes,
memory Inactive anon bytes,
memory Inactive bytes,
memory MemAvailable bytes,
memory MemFree bytes,
memory Shmem bytes.

Most of these metrics are related to memory usage during the execution of the tests.
\subsection{Traces}
Due to the reasons described in Section \ref{sec:desc_traces}, we did not analyze the traces.

\section{Future work based on the dataset}
\label{sec:future}
As this study focuses on software execution anomalies through negative testing, the log files, in particular, can provide a great dataset for log anomaly detection. Most of the datasets used in log anomaly research consist of logs from monolithic systems, and even those from distributed systems simply aggregate everything together \cite{loghub}. The preliminary analysis of LO2 data has shown that it is not possible for our dataset. This raises several challenges: 
\begin{itemize}
    \item If the error is not known, how to select the appropriate service logs for training?
    \item When running models for multiple services, how to set the threshold and normalize the results between services?
    \item How to determine the root cause when a single error causes anomalies on multiple services?
\end{itemize}
Furthermore, the dataset is large enough that each service could be studied separately. In short, the characteristics of the data, as well as its sheer volume, make LO2 a great dataset for future work on log anomaly detection. 

While we did not introduce performance anomalies, the dataset provides various metrics that were collected during runtime. These can be used as a basis for time-series analysis by themselves or by fusion with the logs to improve anomaly detection. However, we expect the impact of test case selection to remain light on most performance metrics as the system under study manages to handle the errors gracefully. 

The long-term goal for future research is to develop a common schema in order to harmonize varied data formats and structures emanating from traces, logs, and metrics. This schema should permit flexible representation of various architectures while preserving semantic consistency; hence, shared identifiers, such as request IDs and timestamps, are included across all data types to allow their correlation for the exploration of inter-dependencies.

Such a dataset would facilitate future advanced anomaly detection on logs, metrics, and traces in microservice systems, and, in particular, it would support fusion methods relying on multiple modalities. Temporal alignment of data can provide researchers with an overview of system behavior over time and help to detect anomalous patterns that span across sources of data.

\section{Contributing to the dataset}
\label{sec:udaptes}
We highlight the following possibilities to contribute to this dataset and create an updated, more fine-grained version:

\textbf{Integrate proper tracing.}
This can be achieved by either instrumenting the traces in the source code or investigating whether \texttt{networknt} own tracing system \footnote{\url{https://doc.networknt.com/architecture/traceability/}} integrates with Light-OAuth2.

\textbf{Change the tasks' spawn rate.} In the current version of the dataset, all used Locust tasks (correct and erroneous) had an equal chance of getting executed.

In Locust, it is possible to provide a relative weight to each task, affecting how likely it is to be picked, as well as modifying the amount of spawned users at runtime, making it possible to create more complicated load scenarios more closely representing a production deployment.

It is possible to, e.g., distribute calls with successful results (response 200 or 302) relative to errors (400, 401, 404, 500) by using some known real-life distribution, like one found in \cite{844498,world_cup_1998, sona_ghahremani_2021_5145855}, or introducing load spikes during the execution of the run.

\textbf{Capture metrics for each service}, and not only host-wide metrics.

\textbf{Introduce host/hardware anomalies} such as slow/breaking connection, slow disk read/write speed, or crashing and redeployment of containers, similar to \cite{yu2023nezha, LeeYCSL23, li2022actionable}.

\textbf{Use a different MSS system} while following similar test generation, execution, and data collection procedures. This will allow to have a dataset consisting of several MSS for comparison, particularly if it is another implementation of the OAuth2.0 protocol.

\textbf{Community contributions.} We encourage the community to contribute datasets improving upon the aforementioned aspects  or providing new systems by opening an issue on our community repository\footnote{\url{https://github.com/M3SOulu/LO2-Dataset-Community}} and filling in the necessary information. Using GitHub issues
allows a public peer review of the candidate dataset. 
Similarly, issues can be used to request a change or fix a bug in the current dataset or its replication framework.

\section{How to use the dataset}
\label{sec:howto}
The dataset is prepared and shared according to FAIR principles \cite{force11_fair} - Findable, Accessible, Interoperable, and Re-usable.

The dataset consists of two repositories shared via Zenodo \cite{replication, dataset_repo}. One of them consists of the replication package \cite{replication} with scripts to deploy the entire system, run the Locust tests, and query the logs, metrics, and traces.

The other one \cite{dataset_repo} contains the full dataset with 1740 runs as well as some cleaned sample data used for the preliminary analysis in this work. A detailed explanation of the file and data structure is available in the dataset description.

We plan to update the dataset to solve some of the issues of Section \ref{sec:udaptes}; so when using the dataset, please cite this paper as well as the exact version of the Zenodo data you used for your study.

\section{License}
\label{sec:license}
We provide our dataset under the CC BY 4.0 license \cite{creative_commons_2024}, allowing unrestricted use, distribution, and adaptation, provided that appropriate credit is given to the original authors.

\section{Threats to Validity}
\label{sec:Threats}
This section presents the possible threats to the validity of our study, grouped according to Construct, Internal, and External validity, as per Wohlin et al.~\cite{wohlin2012experimentation}.

\textbf{Construct Validity} refers to how well our measurements reflect our intended assessment.
Specific design choices, such as how we measure and filter data, may affect our findings.
To address this threat, we relied on previous work \cite{esposito2024validate,falessi2023enhancing1,lenarduzzi2019technical} when we made such decisions, allowing us to be consistent with what already existed.

Light-OAuth2 has incomplete documentation of its APIs, so the construction of Locust tasks is subject to misunderstanding. However, tests were prepared by two authors who reviewed each other's work. Most CRUD operations operation are standard and cover meaningful error cases that were implemented in our tests, while for OAuth2.0 flows, the official standard of OAuth2.0 \cite{rfc6749} serves as an additional source for the construction of Locust tasks, so we could verify that Light-OAuth2.0 implementation follows the standard protocol. 

The issues encountered regarding tracing data made it impossible to provide a dataset complete with three modalities (logs, metrics, and traces). The potential solutions to the tracing issue are: (1) use a different system with properly instrumented tracing; (2) manually instrument tracing in the source code of each service of Light-OAuth2; (3) fix and rely on \texttt{networknt}'s own implementation of tracing. For the first solution, we are not aware of other easily deployable industrial OSS microservice systems, hence the choice of Light-OAuth2; the other solutions require a great deal of manual effort, which would not represent the actual approach to deploying and tracing microservice at scale adopted in the Industry.

\textbf{Internal Validity} concerns how much the cause-effect relationship in our study can be trusted. Factors that might have threatened these results include a lack of standardization in testing conditions and biased test design. For example, tests run under different conditions, and focusing on specific microservices could warp the dataset. We performed extensive testing of CRUD operations for data-managing services and OAuth2.0 flows like Authorization Code with/without PKCE, Client Credentials, and Refresh Token flow, excluding insecure ones such as Resource Owner Password flow\footnote{\url{https://auth0.com/docs/get-started/authentication-and-authorization-flow/resource-owner-password-flow}}. Furthermore, we avoided non-protocol-specific features of the Light-OAuth2 system to ensure portability across OAuth2.0 implementations and running repeated tests on the same machine over an extended period captured load variations, such as during university downtime. We minimized assumptions related to databases by abstracting away from database deployment and using service-managed configurations. Moreover, the database was reset for each test run to prevent performance degradation due to data accumulation. We did not induce hardware anomalies such as network delay or slow disk speed, focusing only on system response to correct and erroneous API use to minimize factors that could be the cause of anomalies in logs or metrics.

\textbf{External Validity} concerns the generalizability of our findings. The dataset, obtained from the MS-based Light-OAuth2 project, is intended to reflect real-world production environments. However, the configurations, choices of databases, and operational loads may limit its 
applicability beyond our controlled setup.
To alleviate these issues, we made sure to use the standard Docker Compose deployment as provided by the authors of the project, only injecting the Jaeger and Prometheus containers, which is standard practice for tracing/metric collection. The choice of the specific database (MySQL) was arbitrary, but it is indeed one of the officially supported databases. The server we could use for the deployment of the system is constrained by resources available at our university and may not represent the capabilities of a real-world production cloud server. 
The selection of Locust tasks at runtime selected each task with equal probability, thus potentially providing an unrealistic distribution of correct vs. erroneous calls to the APIs. However, the Locust tool allows easy modification of task weights as well as changes to the distribution of Users at test runtime, so it should be possible to modify the source code of our tests \cite{replication} to change the distribution of errors.
For example, the distribution of different HTTP error codes can be inferred from datasets such as World Cup 1998 \cite{844498, world_cup_1998, sona_ghahremani_2021_5145855}. However, the weights would need to be computed at runtime using reflection based on the selected tasks, so we opted not to resort to it in this work.
Despite these constraints, the diversity of test scenarios and the volume of collected data provide a robust foundation for capturing varied system behaviors and error conditions. Additionally, data can be sampled from our dataset with the desired frequencies of correct and error tests without generating a new version of the dataset.

\section{Conclusion}
\label{sec:Conclusion}
In conclusion, the aim of this work was to provide a novel multi-modal dataset of monitoring data consisting of logs, metrics, and traces of a production OSS microservice system.

To achieve this goal, 
we deployed the Light-OAuth2 MSS and created and executed Locust tests targeting different APIs of the system. We set up the data collection pipeline to gather container logs from Docker, host metrics by injecting Prometheus, and traces by injecting the Jaeger agent. We provided the dataset of logs and metrics, replication instructions, and some data processing pipelines. We described some initial attempts and conclusions of analyzing logs, the most prominent metrics due to PCA, and challenges in gathering traces for this system.
This dataset can enable work on anomaly detection in logs and metrics of an MSS, particularly in methods attempting a fusion of log and metric modalities.

\section*{Acknowledgement}
This work has been funded by the Research Council of Finland (grants n. 359861 and 349488 - MuFAno) and Business Finland (grant 6GSoft).
\bibliographystyle{ACM-Reference-Format}
\bibliography{main}


\begin{thebibliography}{34}


\ifx \showCODEN    \undefined \def \showCODEN     #1{\unskip}     \fi
\ifx \showISBNx    \undefined \def \showISBNx     #1{\unskip}     \fi
\ifx \showISBNxiii \undefined \def \showISBNxiii  #1{\unskip}     \fi
\ifx \showISSN     \undefined \def \showISSN      #1{\unskip}     \fi
\ifx \showLCCN     \undefined \def \showLCCN      #1{\unskip}     \fi
\ifx \shownote     \undefined \def \shownote      #1{#1}          \fi
\ifx \showarticletitle \undefined \def \showarticletitle #1{#1}   \fi
\ifx \showURL      \undefined \def \showURL       {\relax}        \fi
\providecommand\bibfield[2]{#2}
\providecommand\bibinfo[2]{#2}
\providecommand\natexlab[1]{#1}
\providecommand\showeprint[2][]{arXiv:#2}

\bibitem[Amadini et~al\mbox{.}(2024)]%
        {amadini2024pick}
\bibfield{author}{\bibinfo{person}{Roberto Amadini}, \bibinfo{person}{Simone Gazza}, \bibinfo{person}{Jacopo Soldani}, \bibinfo{person}{Monica Vitali}, \bibinfo{person}{Antonio Brogi}, \bibinfo{person}{Stefano Forti}, \bibinfo{person}{Saverio Giallorenzo}, \bibinfo{person}{Pierluigi Plebani}, \bibinfo{person}{Francisco Ponce}, {and} \bibinfo{person}{Gianluigi Zavattaro}.} \bibinfo{year}{2024}\natexlab{}.
\newblock \showarticletitle{Pick a Flavour: Towards Sustainable Deployment of Cloud-Edge Applications}. In \bibinfo{booktitle}{\emph{International Symposium on Logic-Based Program Synthesis and Transformation}}. Springer, \bibinfo{pages}{117--127}.
\newblock


\bibitem[Arlitt and Jin(2000)]%
        {844498}
\bibfield{author}{\bibinfo{person}{M. Arlitt} {and} \bibinfo{person}{T. Jin}.} \bibinfo{year}{2000}\natexlab{}.
\newblock \showarticletitle{A workload characterization study of the 1998 World Cup Web site}.
\newblock \bibinfo{journal}{\emph{IEEE Network}} \bibinfo{volume}{14}, \bibinfo{number}{3} (\bibinfo{year}{2000}), \bibinfo{pages}{30--37}.
\newblock
\href{https://doi.org/10.1109/65.844498}{doi:\nolinkurl{10.1109/65.844498}}


\bibitem[Bakhtin et~al\mbox{.}(2025)]%
        {bakhtin2025network}
\bibfield{author}{\bibinfo{person}{Alexander Bakhtin}, \bibinfo{person}{Matteo Esposito}, \bibinfo{person}{Valentina Lenarduzzi}, {and} \bibinfo{person}{Davide Taibi}.} \bibinfo{year}{2025}\natexlab{}.
\newblock \showarticletitle{Network Centrality as a New Perspective on Microservice Architecture}. In \bibinfo{booktitle}{\emph{2025 IEEE 22nd International Conference on Software Architecture (ICSA)}}.
\newblock
\href{https://doi.org/10.1109/ICSA65012.2025.00017}{doi:\nolinkurl{10.1109/ICSA65012.2025.00017}}


\bibitem[Bakhtin et~al\mbox{.}(2024a)]%
        {dataset_repo}
\bibfield{author}{\bibinfo{person}{Alexander Bakhtin}, \bibinfo{person}{Jesse Nyyssölä}, \bibinfo{person}{Yuqing Wang}, \bibinfo{person}{Noman Ahmad}, \bibinfo{person}{Ke Ping}, \bibinfo{person}{Matteo Esposito}, \bibinfo{person}{Mika Mäntylä}, {and} \bibinfo{person}{Davide Taibi}.} \bibinfo{year}{2024}\natexlab{a}.
\newblock \bibinfo{title}{LO2: Microservice Dataset of Logs and Metrics [Data set]}.
\newblock
\href{https://doi.org/10.5281/zenodo.14257989}{doi:\nolinkurl{10.5281/zenodo.14257989}}


\bibitem[Bakhtin et~al\mbox{.}(2024b)]%
        {replication}
\bibfield{author}{\bibinfo{person}{Alexander Bakhtin}, \bibinfo{person}{Jesse Nyyssölä}, \bibinfo{person}{Yuqing Wang}, \bibinfo{person}{Noman Ahmad}, \bibinfo{person}{Ke Ping}, \bibinfo{person}{Matteo Esposito}, \bibinfo{person}{Mika Mäntylä}, {and} \bibinfo{person}{Davide Taibi}.} \bibinfo{year}{2024}\natexlab{b}.
\newblock \bibinfo{title}{Replication for "LO2: Microservice Dataset of Logs and Metrics"}.
\newblock
\href{https://doi.org/10.5281/zenodo.14229369}{doi:\nolinkurl{10.5281/zenodo.14229369}}


\bibitem[{Creative Commons}(2024)]%
        {creative_commons_2024}
\bibfield{author}{\bibinfo{person}{{Creative Commons}}.} \bibinfo{year}{2024}\natexlab{}.
\newblock \bibinfo{title}{Creative Commons Attribution 4.0 International Public License}.
\newblock
\urldef\tempurl%
\url{https://creativecommons.org/licenses/by/4.0/}
\showURL{%
\tempurl}
\newblock
\shownote{Accessed: 2024-10-29}.


\bibitem[Esposito and Falessi(2024)]%
        {esposito2024validate}
\bibfield{author}{\bibinfo{person}{Matteo Esposito} {and} \bibinfo{person}{Davide Falessi}.} \bibinfo{year}{2024}\natexlab{}.
\newblock \showarticletitle{VALIDATE: A deep dive into vulnerability prediction datasets}.
\newblock \bibinfo{journal}{\emph{Information and Software Technology}} (\bibinfo{year}{2024}), \bibinfo{pages}{107448}.
\newblock


\bibitem[Falessi et~al\mbox{.}(2023)]%
        {falessi2023enhancing1}
\bibfield{author}{\bibinfo{person}{Davide Falessi}, \bibinfo{person}{Simone~Mesiano Laureani}, \bibinfo{person}{Jonida {\c{C}}arka}, \bibinfo{person}{Matteo Esposito}, {and} \bibinfo{person}{Daniel~Alencar da Costa}.} \bibinfo{year}{2023}\natexlab{}.
\newblock \showarticletitle{Enhancing the defectiveness prediction of methods and classes via {JIT}}.
\newblock \bibinfo{journal}{\emph{Empir. Softw. Eng.}} \bibinfo{volume}{28}, \bibinfo{number}{2} (\bibinfo{year}{2023}), \bibinfo{pages}{37}.
\newblock
\href{https://doi.org/10.1007/S10664-022-10261-Z}{doi:\nolinkurl{10.1007/S10664-022-10261-Z}}


\bibitem[{FORCE11}(2014)]%
        {force11_fair}
\bibfield{author}{\bibinfo{person}{{FORCE11}}.} \bibinfo{year}{2014}\natexlab{}.
\newblock \bibinfo{title}{The {FAIR} Data Principles}.
\newblock
\urldef\tempurl%
\url{https://force11.org/info/the-fair-data-principles/}
\showURL{%
\tempurl}
\newblock
\shownote{Accessed: 2024-12-02}.


\bibitem[{FudanSELab}(2024)]%
        {DeepTraLog}
\bibfield{author}{\bibinfo{person}{{FudanSELab}}.} \bibinfo{year}{2024}\natexlab{}.
\newblock \bibinfo{title}{DeepTraLog}.
\newblock \bibinfo{howpublished}{\url{https://github.com/FudanSELab/DeepTraLog}}.
\newblock


\bibitem[Ghahremani(2021)]%
        {sona_ghahremani_2021_5145855}
\bibfield{author}{\bibinfo{person}{Sona Ghahremani}.} \bibinfo{year}{2021}\natexlab{}.
\newblock \bibinfo{booktitle}{\emph{1998 World Cup Website Access Logs}}.
\newblock
\href{https://doi.org/10.5281/zenodo.5145855}{doi:\nolinkurl{10.5281/zenodo.5145855}}


\bibitem[Hardt(2012)]%
        {rfc6749}
\bibfield{author}{\bibinfo{person}{D. Hardt}.} \bibinfo{year}{2012}\natexlab{}.
\newblock \bibinfo{title}{{The OAuth 2.0 Authorization Framework}}.
\newblock \bibinfo{howpublished}{Request for Comments 6749}.
\newblock
\urldef\tempurl%
\url{https://www.rfc-editor.org/rfc/rfc6749}
\showURL{%
\tempurl}


\bibitem[Hora(2024)]%
        {hora2024test}
\bibfield{author}{\bibinfo{person}{Andre Hora}.} \bibinfo{year}{2024}\natexlab{}.
\newblock \showarticletitle{Test polarity: detecting positive and negative tests}. In \bibinfo{booktitle}{\emph{Companion Proceedings of the 32nd ACM International Conference on the Foundations of Software Engineering}}. \bibinfo{pages}{537--541}.
\newblock


\bibitem[{IntelligentDDS}(2024)]%
        {Nezha}
\bibfield{author}{\bibinfo{person}{{IntelligentDDS}}.} \bibinfo{year}{2024}\natexlab{}.
\newblock \bibinfo{title}{Nezha}.
\newblock \bibinfo{howpublished}{\url{https://github.com/IntelligentDDS/Nezha}}.
\newblock


\bibitem[Landauer et~al\mbox{.}(2024)]%
        {criticalreviewlog}
\bibfield{author}{\bibinfo{person}{Max Landauer}, \bibinfo{person}{Florian Skopik}, {and} \bibinfo{person}{Markus Wurzenberger}.} \bibinfo{year}{2024}\natexlab{}.
\newblock \showarticletitle{A Critical Review of Common Log Data Sets Used for Evaluation of Sequence-Based Anomaly Detection Techniques}.
\newblock \bibinfo{journal}{\emph{Proc. ACM Softw. Eng.}} \bibinfo{volume}{1}, \bibinfo{number}{FSE}, Article \bibinfo{articleno}{61} (\bibinfo{date}{July} \bibinfo{year}{2024}), \bibinfo{numpages}{22}~pages.
\newblock
\href{https://doi.org/10.1145/3660768}{doi:\nolinkurl{10.1145/3660768}}


\bibitem[{Lawrence Berkeley National Laboratory}(1998)]%
        {world_cup_1998}
\bibfield{author}{\bibinfo{person}{{Lawrence Berkeley National Laboratory}}.} \bibinfo{year}{1998}\natexlab{}.
\newblock \bibinfo{title}{1998 World Cup Dataset}.
\newblock
\urldef\tempurl%
\url{https://ita.ee.lbl.gov/html/contrib/WorldCup.html}
\showURL{%
\tempurl}
\newblock
\shownote{Accessed: 2024-10-29}.


\bibitem[Lee et~al\mbox{.}(2023a)]%
        {Eadro}
\bibfield{author}{\bibinfo{person}{Cheryl Lee}, \bibinfo{person}{Tianyi Yang}, \bibinfo{person}{Zhuangbin Chen}, \bibinfo{person}{Yuxin Su}, {and} \bibinfo{person}{Michael~R. Lyu}.} \bibinfo{year}{2023}\natexlab{a}.
\newblock \bibinfo{title}{Eadro}.
\newblock
\href{https://doi.org/10.5281/zenodo.7615394}{doi:\nolinkurl{10.5281/zenodo.7615394}}
\newblock
\shownote{[Dataset]}.


\bibitem[Lee et~al\mbox{.}(2023b)]%
        {LeeYCSL23}
\bibfield{author}{\bibinfo{person}{Cheryl Lee}, \bibinfo{person}{Tianyi Yang}, \bibinfo{person}{Zhuangbin Chen}, \bibinfo{person}{Yuxin Su}, {and} \bibinfo{person}{Michael~R. Lyu}.} \bibinfo{year}{2023}\natexlab{b}.
\newblock \showarticletitle{Eadro: An End-to-End Troubleshooting Framework for Microservices on Multi-source Data}. In \bibinfo{booktitle}{\emph{45th {IEEE/ACM} International Conference on Software Engineering, {ICSE} 2023, Melbourne, Australia, May 14-20, 2023}}. \bibinfo{publisher}{{IEEE}}, \bibinfo{pages}{1750--1762}.
\newblock


\bibitem[Lenarduzzi et~al\mbox{.}(2019)]%
        {lenarduzzi2019technical}
\bibfield{author}{\bibinfo{person}{Valentina Lenarduzzi}, \bibinfo{person}{Nyyti Saarim{\"a}ki}, {and} \bibinfo{person}{Davide Taibi}.} \bibinfo{year}{2019}\natexlab{}.
\newblock \showarticletitle{The Technical Debt Dataset}. In \bibinfo{booktitle}{\emph{Conference on Predictive Models and Data Analytics in Software Engineering}}.
\newblock


\bibitem[Li et~al\mbox{.}(2022)]%
        {li2022actionable}
\bibfield{author}{\bibinfo{person}{Zeyan Li}, \bibinfo{person}{Nengwen Zhao}, \bibinfo{person}{Mingjie Li}, \bibinfo{person}{Xianglin Lu}, \bibinfo{person}{Lixin Wang}, \bibinfo{person}{Dongdong Chang}, \bibinfo{person}{Xiaohui Nie}, \bibinfo{person}{Li Cao}, \bibinfo{person}{Wenchi Zhang}, \bibinfo{person}{Kaixin Sui}, {et~al\mbox{.}}} \bibinfo{year}{2022}\natexlab{}.
\newblock \showarticletitle{Actionable and interpretable fault localization for recurring failures in online service systems}. In \bibinfo{booktitle}{\emph{Proceedings of the 30th ACM Joint European Software Engineering Conference and Symposium on the Foundations of Software Engineering}}. \bibinfo{pages}{996--1008}.
\newblock


\bibitem[M{\"a}ntyl{\"a} et~al\mbox{.}(2024)]%
        {Mantyla_LogLead_-_Fast_2024}
\bibfield{author}{\bibinfo{person}{Mika~V M{\"a}ntyl{\"a}}, \bibinfo{person}{Yuqing Wang}, {and} \bibinfo{person}{Jesse Nyyss{\"o}l{\"a}}.} \bibinfo{year}{2024}\natexlab{}.
\newblock \showarticletitle{Loglead-fast and integrated log loader, enhancer, and anomaly detector}. In \bibinfo{booktitle}{\emph{2024 IEEE International Conference on Software Analysis, Evolution and Reengineering (SANER)}}. IEEE, \bibinfo{pages}{395--399}.
\newblock


\bibitem[NetManAIOps(2022)]%
        {Dejavu}
\bibfield{author}{\bibinfo{person}{NetManAIOps}.} \bibinfo{year}{2022}\natexlab{}.
\newblock \bibinfo{title}{D\'ej\`aVu}.
\newblock
\href{https://doi.org/10.5281/zenodo.6955909}{doi:\nolinkurl{10.5281/zenodo.6955909}}
\newblock
\shownote{[Dataset]}.


\bibitem[Nyyssölä and Mäntylä(2024)]%
        {nyyssola2024unsuper}
\bibfield{author}{\bibinfo{person}{Jesse Nyyssölä} {and} \bibinfo{person}{Mika Mäntylä}.} \bibinfo{year}{2024}\natexlab{}.
\newblock \showarticletitle{Speed and Performance of Parserless and Unsupervised Anomaly Detection Methods on Software Logs}. In \bibinfo{booktitle}{\emph{2024 IEEE 24th International Conference on Software Quality, Reliability and Security (QRS)}}. \bibinfo{pages}{657--666}.
\newblock
\href{https://doi.org/10.1109/QRS62785.2024.00071}{doi:\nolinkurl{10.1109/QRS62785.2024.00071}}


\bibitem[Owotogbe et~al\mbox{.}(2024)]%
        {chaosengineering2024}
\bibfield{author}{\bibinfo{person}{Joshua Owotogbe}, \bibinfo{person}{Indika Kumara}, \bibinfo{person}{Willem-Jan Van~Den Heuvel}, {and} \bibinfo{person}{Damian~Andrew Tamburri}.} \bibinfo{year}{2024}\natexlab{}.
\newblock \bibinfo{title}{Chaos Engineering: A Multi-Vocal Literature Review}.
\newblock
\showeprint[arxiv]{2412.01416}~[cs.SE]
\urldef\tempurl%
\url{https://arxiv.org/abs/2412.01416}
\showURL{%
\tempurl}


\bibitem[Pham(2024)]%
        {rcaeval}
\bibfield{author}{\bibinfo{person}{Luan Pham}.} \bibinfo{year}{2024}\natexlab{}.
\newblock \bibinfo{title}{RCAEval: A Benchmark for Root Cause Analysis of Microservice Systems}.
\newblock
\href{https://doi.org/10.5281/ZENODO.14590730}{doi:\nolinkurl{10.5281/ZENODO.14590730}}


\bibitem[Schneider and Scandariato(2023)]%
        {schneider2023automatic}
\bibfield{author}{\bibinfo{person}{Simon Schneider} {and} \bibinfo{person}{Riccardo Scandariato}.} \bibinfo{year}{2023}\natexlab{}.
\newblock \showarticletitle{Automatic extraction of security-rich dataflow diagrams for microservice applications written in Java}.
\newblock \bibinfo{journal}{\emph{Journal of Systems and Software}}  \bibinfo{volume}{202} (\bibinfo{year}{2023}), \bibinfo{pages}{111722}.
\newblock


\bibitem[Souppaya et~al\mbox{.}(2017)]%
        {nist800-190}
\bibfield{author}{\bibinfo{person}{Murugiah Souppaya}, \bibinfo{person}{John Morello}, {and} \bibinfo{person}{Karen Scarfone}.} \bibinfo{year}{2017}\natexlab{}.
\newblock \bibinfo{booktitle}{\emph{Application Container Security Guide}}.
\newblock \bibinfo{type}{{T}echnical {R}eport} SP 800-190. \bibinfo{institution}{National Institute of Standards and Technology (NIST)}.
\newblock
\urldef\tempurl%
\url{https://nvlpubs.nist.gov/nistpubs/SpecialPublications/NIST.SP.800-190.pdf}
\showURL{%
\tempurl}


\bibitem[Taibi et~al\mbox{.}(2017)]%
        {taibi2017processes}
\bibfield{author}{\bibinfo{person}{Davide Taibi}, \bibinfo{person}{Valentina Lenarduzzi}, {and} \bibinfo{person}{Claus Pahl}.} \bibinfo{year}{2017}\natexlab{}.
\newblock \showarticletitle{Processes, motivations, and issues for migrating to microservices architectures: An empirical investigation}.
\newblock \bibinfo{journal}{\emph{IEEE Cloud Computing}} \bibinfo{volume}{4}, \bibinfo{number}{5} (\bibinfo{year}{2017}), \bibinfo{pages}{22--32}.
\newblock


\bibitem[{Tsinghua University}(2021)]%
        {AIOpsChallenge2021}
\bibfield{author}{\bibinfo{person}{{Tsinghua University}}.} \bibinfo{year}{2021}\natexlab{}.
\newblock \bibinfo{title}{AIOps Challenge 2021}.
\newblock \bibinfo{howpublished}{\url{https://www.aiops.cn/gitlab/aiops-nankai/data/trace/aiops2021 }}.
\newblock


\bibitem[Wohlin et~al\mbox{.}(2012)]%
        {wohlin2012experimentation}
\bibfield{author}{\bibinfo{person}{Claes Wohlin}, \bibinfo{person}{Per Runeson}, \bibinfo{person}{Martin H{\"{o}}st}, \bibinfo{person}{Magnus~C. Ohlsson}, {and} \bibinfo{person}{Bj{\"{o}}rn Regnell}.} \bibinfo{year}{2012}\natexlab{}.
\newblock \bibinfo{booktitle}{\emph{Experimentation in Software Engineering}}.
\newblock \bibinfo{publisher}{Springer}.
\newblock
\showISBNx{978-3-642-29043-5}


\bibitem[Yu et~al\mbox{.}(2023)]%
        {yu2023nezha}
\bibfield{author}{\bibinfo{person}{Guangba Yu}, \bibinfo{person}{Pengfei Chen}, \bibinfo{person}{Yufeng Li}, \bibinfo{person}{Hongyang Chen}, \bibinfo{person}{Xiaoyun Li}, {and} \bibinfo{person}{Zibin Zheng}.} \bibinfo{year}{2023}\natexlab{}.
\newblock \showarticletitle{Nezha: Interpretable fine-grained root causes analysis for microservices on multi-modal observability data}. In \bibinfo{booktitle}{\emph{Proceedings of the 31st ACM Joint European Software Engineering Conference and Symposium on the Foundations of Software Engineering}}. \bibinfo{pages}{553--565}.
\newblock


\bibitem[Zhang et~al\mbox{.}(2022)]%
        {zhang2022deeptralog}
\bibfield{author}{\bibinfo{person}{Chenxi Zhang}, \bibinfo{person}{Xin Peng}, \bibinfo{person}{Chaofeng Sha}, \bibinfo{person}{Ke Zhang}, \bibinfo{person}{Zhenqing Fu}, \bibinfo{person}{Xiya Wu}, \bibinfo{person}{Qingwei Lin}, {and} \bibinfo{person}{Dongmei Zhang}.} \bibinfo{year}{2022}\natexlab{}.
\newblock \showarticletitle{Deeptralog: Trace-log combined microservice anomaly detection through graph-based deep learning}. In \bibinfo{booktitle}{\emph{Proceedings of the 44th international conference on software engineering}}. \bibinfo{pages}{623--634}.
\newblock


\bibitem[Zhu et~al\mbox{.}(2023)]%
        {loghub}
\bibfield{author}{\bibinfo{person}{Jieming Zhu}, \bibinfo{person}{Shilin He}, \bibinfo{person}{Pinjia He}, \bibinfo{person}{Jinyang Liu}, {and} \bibinfo{person}{Michael~R. Lyu}.} \bibinfo{year}{2023}\natexlab{}.
\newblock \showarticletitle{Loghub: A Large Collection of System Log Datasets for AI-driven Log Analytics}. In \bibinfo{booktitle}{\emph{2023 IEEE 34th International Symposium on Software Reliability Engineering (ISSRE)}}. \bibinfo{pages}{355--366}.
\newblock
\href{https://doi.org/10.1109/ISSRE59848.2023.00071}{doi:\nolinkurl{10.1109/ISSRE59848.2023.00071}}


\bibitem[Zuo et~al\mbox{.}(2020)]%
        {8957683}
\bibfield{author}{\bibinfo{person}{Yuan Zuo}, \bibinfo{person}{Yulei Wu}, \bibinfo{person}{Geyong Min}, \bibinfo{person}{Chengqiang Huang}, {and} \bibinfo{person}{Ke Pei}.} \bibinfo{year}{2020}\natexlab{}.
\newblock \showarticletitle{An Intelligent Anomaly Detection Scheme for Micro-Services Architectures With Temporal and Spatial Data Analysis}.
\newblock \bibinfo{journal}{\emph{IEEE Transactions on Cognitive Communications and Networking}} \bibinfo{volume}{6}, \bibinfo{number}{2} (\bibinfo{year}{2020}), \bibinfo{pages}{548--561}.
\newblock
\href{https://doi.org/10.1109/TCCN.2020.2966615}{doi:\nolinkurl{10.1109/TCCN.2020.2966615}}


\end{thebibliography}

\begin{table*}[]
\centering
\small
\caption{Anomaly Detection F1 Score with DecisionTree for a Given Error Test and Service Log}
\label{tab:log2}
\begin{tabular}{lrrrrrrr}
\multirow{2}{*}{\textbf{Error Test}} & \multicolumn{7}{c}{\textbf{Service Log}} \\ \cmidrule(lr){2-8}
        & \textbf{client} & \textbf{code} & \textbf{key} & \textbf{refresh} & \textbf{service} & \textbf{token} & \textbf{user} \\ \hline
\rowcolor{gray!15}
register\_client\_400\_clientProfile               & \textbf{1.000}          & 0.500          & 0.227          & 0.583          & 0.673          & 0.981          & 0.487          \\
\rowcolor{gray!15}
register\_client\_400\_clientType                  & \textbf{1.000}          & 0.353          & 0.514          & 0.871          & 0.811          & 0.979          & 0.644          \\
\rowcolor{gray!15}
register\_client\_404\_no\_user                    & \textbf{1.000}          & 0.305          & 0.438          & 0.546          & 0.846          & \textbf{1.000}          & 0.640          \\
\rowcolor{gray!15}
delete\_client\_404\_no\_client                    & \textbf{1.000}          & 0.472          & 0.394          & 0.791          & 0.822          & \textbf{1.000}          & 0.685          \\
\rowcolor{gray!15}
get\_client\_404\_no\_client                       & \textbf{1.000}          & 0.772          & 0.372          & 0.419          & 0.765          & 0.959          & 0.510          \\
\rowcolor{gray!15}
get\_client\_page\_400\_no\_page                   & \textbf{1.000}          & 0.328          & 0.596          & 0.400          & 0.746          & 0.971          & 0.585          \\
\rowcolor{gray!15}
update\_client\_400\_clientProfile                 & \textbf{1.000}          & 0.692          & 0.279          & 0.583          & 0.710          & 0.980          & 0.529          \\
\rowcolor{gray!15}
update\_client\_400\_clientType                    & \textbf{1.000}          & 0.723          & 0.354          & 0.769          & 0.706          & 0.991          & 0.396          \\
\rowcolor{gray!15}
update\_client\_404\_clientId                      & \textbf{0.991}          & 0.691          & 0.606          & 0.406          & 0.783          & 0.990          & 0.500          \\
\rowcolor{gray!15}
update\_client\_404\_ownerId                       & \textbf{1.000}          & 0.140          & 0.455          & 0.622          & 0.681          & 0.991          & 0.535          \\
authorization\_code\_client\_id\_missing\_400      & 0.981          & \textbf{1.000}          & 0.034          & 0.349          & 0.783          & 0.957          & 0.641          \\
authorization\_code\_invalid\_password\_401        & 0.900          & \textbf{1.000}          & 0.421          & 0.345          & 0.932          & 0.973          & 0.661          \\
authorization\_code\_invalid\_client\_id\_404      & 0.915          & \textbf{0.980}          & 0.282          & 0.727          & 0.980          & 0.980          & 0.688          \\
authorization\_code\_missing\_response\_type\_400  & 0.972          & \textbf{1.000}          & 0.314          & 0.361          & 0.813          & 0.980          & 0.628          \\
authorization\_code\_response\_not\_code\_400      & 0.861          & \textbf{0.981}          & 0.274          & 0.394          & 0.869          & 0.990          & 0.566          \\
code\_challenge\_invalid\_format\_pkce\_400        & 0.969          & \textbf{1.000}          & 0.286          & 0.418          & 0.774          & \textbf{1.000}          & 0.560          \\
code\_challenge\_too\_long\_pkce\_400              & 0.990          & \textbf{1.000}          & 0.619          & 0.587          & 0.819          & 0.957          & 0.706          \\
code\_challenge\_too\_short\_pkce\_400             & 0.892          & 0.271          & 0.074          & 0.632          & 0.825          & \textbf{1.000}          & 0.519          \\
code\_verifier\_missing\_pkce\_400                 & 0.959          & 0.451          & 0.077          & 0.564          & 0.805          & \textbf{0.967}          & 0.483          \\
code\_verifier\_too\_long\_pkce\_400               & 0.968          & 0.240          & 0.100          & 0.575          & 0.829          & \textbf{1.000}          & 0.624          \\
code\_verifier\_too\_short\_pkce\_400              & \textbf{0.950}          & 0.037          & 0.506          & 0.537          & 0.871          & \textbf{0.950}          & 0.596          \\
invalid\_code\_challenge\_method\_pkce\_400        & 0.983          & \textbf{1.000}          & 0.328          & 0.479          & 0.832          & 0.957          & 0.617          \\
invalid\_code\_verifier\_format\_PKCE\_400         & \textbf{0.920}          & 0.000          & 0.346          & 0.308          & 0.793          & 0.979          & 0.509          \\
\rowcolor{gray!15}
delete\_service\_404\_no\_service                  & 0.945          & 0.603          & 0.635          & 0.692          & \textbf{1.000}          & \textbf{1.000}          & 0.577          \\
\rowcolor{gray!15}
get\_service\_404\_no\_service                     & 0.649          & 0.709          & 0.558          & 0.469          & \textbf{1.000}          & \textbf{1.000}          & 0.500          \\
\rowcolor{gray!15}
get\_service\_page\_400\_no\_page                  & 0.917          & 0.684          & 0.353          & 0.442          & \textbf{1.000}          & 0.972          & 0.505          \\
\rowcolor{gray!15}
register\_service\_400\_service\_id                & \textbf{1.000}          & 0.175          & 0.419          & 0.486          & \textbf{1.000}          & 0.971          & 0.430          \\
\rowcolor{gray!15}
register\_service\_400\_service\_type              & 0.991          & 0.613          & 0.080          & 0.781          & \textbf{1.000}          & 0.963          & 0.680          \\
\rowcolor{gray!15}
register\_service\_404\_no\_user                   & \textbf{0.980}          & 0.207          & 0.453          & 0.448          & 0.939          & \textbf{0.980}          & 0.420          \\
\rowcolor{gray!15}
update\_service\_404\_service\_id                  & 0.959          & 0.196          & 0.100          & 0.345          & \textbf{1.000}          & 0.990          & 0.491          \\
\rowcolor{gray!15}
update\_service\_404\_user\_id                     & 0.944          & 0.507          & 0.619          & 0.493          & \textbf{1.000}          & 0.952          & 0.592          \\
access\_token\_authorization\_form\_401            & 0.980          & 0.419          & 0.560          & 0.718          & 0.943          & \textbf{1.000}          & 0.701          \\
access\_token\_auth\_header\_error\_401            & 0.952          & 0.000          & 0.518          & 0.556          & 0.967          & \textbf{0.991}          & 0.598          \\
access\_token\_client\_id\_not\_found\_404         & \textbf{1.000}          & 0.151          & 0.185          & 0.476          & 0.932          & \textbf{1.000}          & 0.695          \\
access\_token\_client\_secret\_wrong\_401          & \textbf{0.990}          & 0.118          & 0.667          & 0.472          & 0.929          & \textbf{0.990}          & 0.689          \\
access\_token\_form\_urlencoded\_400               & 0.971          & 0.419          & 0.355          & 0.737          & 0.961          & \textbf{1.000}          & 0.614          \\
access\_token\_illegal\_grant\_type\_400           & 0.848          & 0.429          & 0.226          & 0.689          & 0.963          & \textbf{1.000}          & 0.611          \\
access\_token\_missing\_authorization\_header\_400 & 0.957          & 0.157          & 0.513          & 0.358          & 0.957          & \textbf{1.000}          & 0.615          \\
delete\_token\_404                                 & 0.970          & 0.699          & 0.310          & \textbf{1.000}          & 0.810          & 0.960          & 0.531          \\
get\_token\_404                                    & 0.970          & 0.781          & 0.347          & \textbf{1.000}          & 0.771          & 0.974          & 0.551          \\
get\_token\_page\_400\_no\_page                    & 0.889          & 0.271          & 0.279          & \textbf{1.000}          & 0.752          & 0.961          & 0.444          \\
verification\_failed\_pkce\_400                    & \textbf{0.991}          & 0.435          & 0.170          & 0.324          & 0.844          & 0.990          & 0.638          \\ 
\rowcolor{gray!15}
delete\_user\_404\_no\_user                        & 0.970          & 0.675          & 0.226          & 0.640          & 0.717          & 0.954          & \textbf{1.000}          \\
\rowcolor{gray!15}
get\_user\_404\_no\_user                           & 0.863          & 0.138          & 0.345          & 0.318          & 0.757          & 0.935          & \textbf{1.000}          \\
\rowcolor{gray!15}
get\_user\_page\_400\_no\_page                     & 0.967          & 0.713          & 0.512          & 0.281          & 0.736          & 0.980          & \textbf{0.981}          \\
\rowcolor{gray!15}
register\_user\_400\_email\_exists                 & 0.979          & 0.192          & 0.306          & 0.508          & 0.711          & 0.989          & \textbf{1.000}          \\
\rowcolor{gray!15}
register\_user\_400\_no\_password                  & 0.907          & 0.271          & 0.237          & 0.371          & 0.825          & 0.990          & \textbf{1.000}          \\
\rowcolor{gray!15}
register\_user\_400\_password\_no\_match           & 0.871          & 0.441          & 0.385          & 0.535          & 0.765          & \textbf{1.000}          & \textbf{1.000}          \\
\rowcolor{gray!15}
register\_user\_400\_user\_exists                  & 0.973          & 0.073          & 0.103          & 0.508          & 0.830          & 0.951          & \textbf{1.000}          \\
\rowcolor{gray!15}
update\_password\_400\_not\_match                  & 0.915          & 0.318          & 0.207          & 0.700          & 0.729          & 0.971          & \textbf{1.000}          \\
\rowcolor{gray!15}
update\_password\_401\_wrong\_password             & 0.969          & 0.703          & 0.400          & 0.375          & 0.700          & 0.979          & \textbf{1.000}          \\
\rowcolor{gray!15}
update\_password\_404\_user\_not\_found            & 0.860          & 0.085          & 0.250          & 0.539          & 0.794          & 0.981          & \textbf{1.000}          \\
\rowcolor{gray!15}
update\_user\_404\_no\_user                        & 0.957          & 0.105          & 0.494          & 0.387          & 0.831          & 0.957          & \textbf{0.991}          \\  \hline
\textbf{Average}                                   & 0.951 & 0.476 & 0.353 & 0.545 & 0.842 & 0.980 & 0.663 \\
\textbf{Median}                                    & 0.970 & 0.435 & 0.347 & 0.508 & 0.825 & 0.980 & 0.615 \\
\textbf{Min}                                       & 0.649 & 0.000 & 0.034 & 0.281 & 0.673 & 0.935 & 0.396 \\
\textbf{Max}                                       & 1.000 & 1.000 & 0.667 & 1.000 & 1.000 & 1.000 & 1.000
\end{tabular}

\end{table*}

\renewcommand{\tablename}{Listing} 
\setcounter{table}{0}

\begin{table*}[]
    \centering
    \caption{Sample of the Collected Log Data} 
\label{tab:sample_logs}
\begin{tabular}{|p{\linewidth}|} 
\hline
16:29:36.735 [XNIO-1 task-1]  dfzUR0LpRACkFzgALOYflA DEBUG c.n.openapi.ApiNormalisedPath \textless init\textgreater{} - normalised = /oauth2/user \\
16:29:36.736 [XNIO-1 task-1]  dfzUR0LpRACkFzgALOYflA DEBUG c.n.openapi.ApiNormalisedPath \textless init\textgreater{} - normalised = /oauth2/user \\
16:29:36.775 [XNIO-1 task-1]  dfzUR0LpRACkFzgALOYflA INFO  com.networknt.config.Config getConfigStream - Config loaded from default folder for mask.yml \\
16:29:36.777 [XNIO-1 task-1]  dfzUR0LpRACkFzgALOYflA DEBUG c.n.openapi.ApiNormalisedPath \textless init\textgreater{} - normalised = /oauth2/user \\
16:29:36.846 [XNIO-1 task-1]  dfzUR0LpRACkFzgALOYflA DEBUG com.networknt.schema.TypeValidator debug - validate( \textquotedbl 1\textquotedbl , \textquotedbl 1\textquotedbl , page) \\
16:29:36.986 [hz.\_hzInstance\_1\_dev.cached.thread-16]   DEBUG com.zaxxer.hikari.HikariConfig logConfiguration - allowPoolSuspension\dots{}false \\
16:29:36.987 [hz.\_hzInstance\_1\_dev.cached.thread-16]   DEBUG com.zaxxer.hikari.HikariConfig logConfiguration - catalog\dots{}none \\
16:29:36.987 [hz.\_hzInstance\_1\_dev.cached.thread-16]   DEBUG com.zaxxer.hikari.HikariConfig logConfiguration - connectionTestQuery\dots{}none \\
16:29:36.987 [hz.\_hzInstance\_1\_dev.cached.thread-16]   DEBUG com.zaxxer.hikari.HikariConfig logConfiguration - dataSource\dots{}none \\
16:29:36.987 [hz.\_hzInstance\_1\_dev.cached.thread-16]   DEBUG com.zaxxer.hikari.HikariConfig logConfiguration - dataSourceJNDI\dots{}none \\
\multicolumn{1}{|c|}{$\vdots$}\\
\hline
\end{tabular}
\end{table*}
\begin{table*}[]
\caption{Sample of the Collected Metric Data Converted to CSV}
\label{tab:sample_metrics}
\resizebox{\linewidth}{!}{%
\begin{tabular}{lllllll}
timestamp  & test\_name                           & gc\_duration& gc\_duration\_count & gc\_duration\_sum & goroutines &$\cdots$\\
\hline
1719592192 & correct                              & 4.45e-05                              & 8                                & 0.000278365                    & 7       &       \\
1719592202 & correct                              & 6.0054e-05                            & 10                               & 0.000369997                    & 7     &         \\
1719592212 & correct                              & 6.0054e-05                            & 13                               & 0.000462525                    & 7       &       \\
1719592222 & correct                              & 6.0054e-05                            & 16                               & 0.00055404                     & 7        &      \\
1719592232 & correct                              & 6.0054e-05                            & 18                               & 0.000641278                    & 7         &     \\
1719592235 & register\_client\_400\_clientType    & 6.0054e-05                            & 19                               & 0.000670362                    & 7         &     \\
1719592245 & register\_client\_400\_clientType    & 6.0054e-05                            & 21                               & 0.000738679                    & 7       &       \\
1719592248 & register\_client\_400\_clientProfile & 6.0054e-05                            & 21                               & 0.000738679                    & 7       &       \\
1719592253 & register\_client\_400\_clientProfile & 6.0054e-05                            & 23                               & 0.000796302                    & 7      &        \\
1719592266 & register\_client\_404\_no\_user      & 6.0054e-05                            & 26                               & 0.000895424                    & 7       &       \\
1719592271 & register\_client\_404\_no\_user      & 6.0054e-05                            & 28                               & 0.000947592                    & 7       &       \\
1719592274 & update\_client\_400\_clientType      & 6.0054e-05                            & 29                               & 0.000978015                    & 7         &     \\
\multicolumn{1}{c}{$\vdots$} &&&&&&
\end{tabular}%
}
\end{table*}
\begin{table*}[]
\caption{Sample of the Collected Tracing Data}
\label{tab:sample_traces}
\begin{tabular}{lllll}
Trace ID                         & Span ID & Operation Name & Start Time &$\cdots$ \\
\hline
d35bfc89f74f989bc2e291cd2aa0c1eb & b5788b25f6e4db93                & SELECT oauth2.refresh\_token           & 2024-06-28T16:30:32.811058+00:00 &  \\
19f14689f4137c3487f8466ba1fb0ac4 & 80d65e80b0665c02                & GET                                    & 2024-06-28T16:30:32.811603+00:00  &  \\
595970b32aea118a69e797ffc909378f & a623b7ba3b40f6b8                & INSERT oauth2.refresh\_token           & 2024-06-28T16:30:32.811670+00:00  &  \\
859c0ad89a079b2138e47cf9cbad967a & e3da089434118d63                & GET                                    & 2024-06-28T16:30:32.819811+00:00   & \\
4ef7246eb4a2b37eeba91696df2d0076 & c944c05fc1cb2524                & GET                                    & 2024-06-28T16:30:32.842286+00:00   & \\
1f240bcab152c8298aaf3c94b6ea078f & 7048b2701a4e172a                & SELECT oauth2.refresh\_token           & 2024-06-28T16:30:32.854248+00:00   & \\
8f69a47b816d6b0679fb76f0a3a80413 & 7cb1329e9c49492d                & INSERT oauth2.refresh\_token           & 2024-06-28T16:30:32.855017+00:00  &  \\
4bdde28b2f37b8a3bbd581df74562094 & 4108c46da844af4d                & GET                                    & 2024-06-28T16:30:32.869853+00:00  &  \\
dc07f0163bcaa297d38663e690421fa0 & 56f71e99f0de8d81                & DELETE oauth2.refresh\_token           & 2024-06-28T16:30:32.870173+00:00  &  \\
9aee7bca43b4cb46890c3d859cfff88d & 83d608d58133c299                & GET                                    & 2024-06-28T16:30:32.900300+00:00  &  \\
d3b7e505a5acd84ed23a044472948c8d & 0edd310418ed7138                & SELECT oauth2.refresh\_token           & 2024-06-28T16:30:32.930627+00:00  &  \\
4c1c8d9af718143ade073b3366b84b82 & df9e38a001f2f0e0                & SELECT oauth2.refresh\_token           & 2024-06-28T16:30:32.931304+00:00  &  \\
ac725e773cad543871b931a94630614c & 491f90da227afd4e                & DELETE oauth2.refresh\_token           & 2024-06-28T16:30:32.953960+00:00  &  \\
c4db9e752ace6803015728138095ecb3 & 8643743bdceafd0b                & GET                                    & 2024-06-28T16:30:32.985495+00:00   & \\
\multicolumn{1}{c}{$\vdots$}   & & &   &
\end{tabular}
\end{table*}

\end{document}